\documentclass[twocolumn,pra,aps]{revtex4}
\usepackage[T1]{fontenc}
\usepackage[latin9]{inputenc}
\usepackage{color}
\usepackage{array}
\usepackage{amsmath}
\usepackage{stackrel}
\usepackage{graphicx}
\usepackage{appendix}
\newcommand{\ket}[1]{\vert{#1}\rangle}
\usepackage{appendix}
\usepackage{hyperref}
\usepackage{color}
\definecolor{med-blue}{RGB}{25,25,112} 
\hypersetup{colorlinks, linkcolor={red},citecolor={blue}, urlcolor={blue}}
\usepackage{natbib}
\allowdisplaybreaks

\providecommand{\tabularnewline}{\\}

\begin{document}

\title{Experimental realization of nondestructive discrimination of Bell
states using a five-qubit quantum computer}
\author{Mitali Sisodia, Abhishek Shukla and Anirban Pathak}
\email{anirban.pathak@gmail.com}
\affiliation{Jaypee Institute of Information Technology, A 10, Sector 62, Noida,
UP 201307, India}

\begin{abstract}
A scheme for distributed quantum measurement that allows nondestructive
or indirect Bell measurement was proposed by Gupta et al., (Int. J.
Quant. Infor. \textbf{5} (2007) 627) and subsequently realized experimentally
using an NMR-based three-qubit quantum computer by Samal et al., (J.
Phys. B, \textbf{43} (2010) 095508). In the present work, a similar
experiment is performed using the five-qubit super-conductivity-based
quantum computer, which has been recently placed in cloud by IBM Corporation.
The experiment confirmed that the Bell state can be constructed and
measured in a nondestructive manner with a reasonably high fidelity. A comparison
of the outcomes of this study and the results obtained earlier in the
NMR-based experiment has also been performed. The study indicates that to make a scalable SQUID-based computer, errors by the gates (in the present technology) have to be reduced considerably.
\end{abstract}
\maketitle


\section{Introduction}

Discrimination of orthogonal entangled states play a very crucial
role in quantum information processing. There exist various alternative
proposals for the same (see \cite{gupta2007general,NDBSD,munro2005efficient,wang2013nondestructive,zheng2016complete}
and references therein). A particularly important variant of state
discrimination schemes is, non-destructive discrimination of entangled
states \cite{gupta2007general,NDBSD,munro2005efficient,wang2013nondestructive,zheng2016complete},
in which the state is not directly measured. The measurement is performed
over some ancilla qubits/qudits and the original state remains unchanged.
Proposals for such non-destructive measurements in optical quantum
information processing using Kerr type nonlinearity have been discussed
in Refs. \cite{munro2005efficient,li2000non,wang2013nondestructive} and
references therein. Such optical schemes of nondestructive discrimination
are important as they are frequently used in designing entanglement
concentration protocols \cite{Banerjee2015}. In the similar line, some
of the present authors proposed an interesting scheme for generalized
orthonormal qudit Bell state discrimination \cite{NDBSD}
and provided an explicit quantum circuit for the task. In Ref. \cite{NDBSD},
it was also established that the use of distributed measurement (i.e.,
non-destructive measurement where the measurement task is distributed
or outsourced to ancilla qubits) have useful applications in reducing
quantum communication complexity under certain conditions. The work
was further generalized in \cite{gupta2007general}, where the relevance
of the quantum circuits for nondestructive discrimination was established
in the context of quantum error correction, measurement-based quantum
computation, Bell state discrimination across a quantum network involving
multiple parties, and optimization of the quantum communication complexity
for performing measurements in distributed quantum computing. Subsequently,
applications of the non-destructive discrimination of entangled states
have been proposed in various other works, too. Specifically, in \cite{jain2009secure},
a scheme for two-way secure direct quantum communication, which was
referred to as quantum conversation, was developed using the nondestructive
discrimination scheme. In a more general scenario, Luo et al., proposed
a scheme for multi-party quantum private comparison based on $d$-dimensional
entangled states \cite{luo2014multi}, where all the participants are required
to perform non-destructive measurement. These applications and the
fact that the scheme proposed in \cite{NDBSD} has been
experimentally implemented for Bell state discrimination using an
NMR-based 3-qubit quantum computer have motivated us to perform nondestructive
Bell state discrimination using another experimental platform. Specifically,
in this work we aim to realize nondestructive Bell state discrimination
using a 5-qubit superconductivity- (SQUID)-based quantum computer \cite
{IBMQE, devitt2016performing}, which has been recently  placed in cloud by IBM Corporation. 
\footnote{Interested readers may find a detailed user 
guide on how to use this computer at \cite
{IBMQE}, and a lucid description of the working principle of a superconductivity-based quantum computer in Ref. \cite{steffen2011quantum}}. This quantum
computer was placed in the cloud in 2016. It immediately drew considerable
attention of the quantum information processing community, and several
quantum information tasks have already been performed using this quantum
computer on cloud.  Specifically, in the domain of quantum communication, experimental realization of teleoprtation of single-qubit 
quantum state \cite{fedortchenko2016quantum} and two-qubit quantum state using optimal resources \cite{sisodia2017design}
have been reported; in the field of quantum foundation, violation of  multi-partite Mermin inequality has been realized for 3, 4\color{red},\color{black} and 5 parties \cite{alsina2016experimental}; an information theoretic version of uncertainty and measurement 
reversibility has also been implemented \cite{berta2016entropic}; in the area of quantum computation, a comparison of two architectures using demonstration of an algorithm has also been performed \cite{linke2017experimental}.  Thus,
we can see that IBM quantum computer has already been successfully used to
realize various tasks belonging to different domains of quantum information processing. However, to the best of our knowledge, IBM quantum computer is not yet used to perform nondestructive discrimination of the orthogonal entangled states, and the performance
of IBM quantum computer is not yet properly compared with the performance of
the quantum computers implemented using liquid NMR technology. In the
present work, we aim to perform such a comparison, subject to a specific
task. To be precise, we wish to compare the performances of IBM quantum
computer and an NMR-based quantum computer with respect to the nondestructive
discrimination of Bell states. There is another important reason for
testing fundamentally important quantum circuits (in our case, quantum circuit for nondestructive discrimination of Bell states)  using the IBM quantum computer
and/or a similar platform- it is now understood that liquid NMR-based
technology is not scalable, and it will not lead to a real scalable
quantum computer. However, it is widely believed that a SQUID-based quantum computer can be made scalable in future. In fact, an
ideal quantum information processor should satisfy Di-Vincenzo's criteria
\cite{divincenzo1997topics}. One
of these criteria requires the  realization of a large quantum register, which
is still a prime technological challenge for experimental quantum
information processing \cite{cory1997ensemble}.
Although NMR-based realizations cannot be scaled, a ray of hope
is generated in the recent past after the introduction of the  relatively new architectures, like solid-state spin systems (nitrogen
vacancy centers in diamond and phosphorous vacancy centers in silicon)
and superconducting-qubits based systems \cite{cottet2002implementation} that
have the potential to become scalable. Among these technologies, owing
to the scalability and functionality of superconducting-qubit registers, they
have emerged as the best candidate for quantum information processing.
Currently, various types of basic superconducting-qubits, namely
Josephson-junction qubits, Phase qubits, Transmon qubits and Potential qubits  are used \cite{clarke2008superconducting,wendin2007quantum}.
SQUID-based quantum information processing architectures have not
only attained the popularity among the researchers, but have also  led
to the path for commercialization of quantum computers. Although a
universal quantum computer with large qubit register is still a distant
hope, large qubit registers to perform specific tasks have been devised.
For example, a quantum computer with register size of 512 qubits was
sold by D-Wave to Google and NASA \cite{dwavenews}.
The machine has been deployed to tackle classification problems that are 
useful in image-recognition and has outperformed its classical counterpart \cite{QCvsCC}.
 As mentioned above, the potential scalability of the
SQUID-based systems  has also motivated us to perform experimental
realization of the Bell state discrimination circuit using a SQUID-based
5 qubit quantum computer. 

Rest of the paper is organized as follows. In Sec. \ref{sec:Method},
we have described the quantum circuits (both theoretical and experimental)
used here to perform Bell state discrimination and the method adopted
here to perform quantum state tomography. In Sec. \ref{sec:Results}, the results
of the experimental realization of the circuits described in the previous
section are reported and analyzed. Finally, the paper is concluded
in Sec. \ref{sec:Conclusion}

\section{Quantum circuits and method used for nondestructive discrimination
of Bell states \label{sec:Method}}

\begin{figure}
\begin{centering}
\includegraphics[width=8cm]{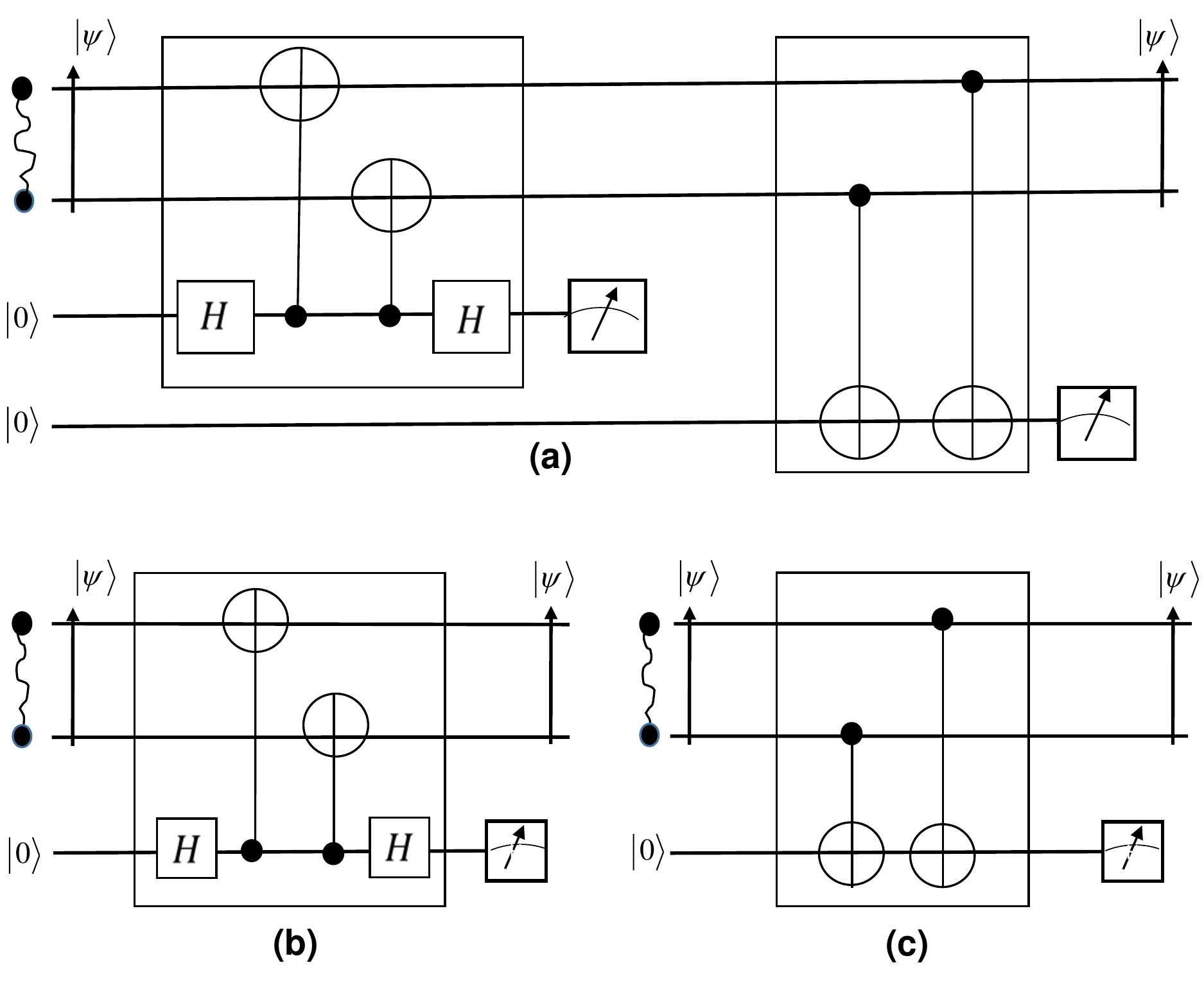}
\par\end{centering}
\caption{Circuits for non-destructive discrimination of Bell-state  \cite{NDBSD}. (a) A four qubit circuit (two system-qubits and two ancilla-qubits) which discriminates all four Bell-states. System-qubits are used for Bell-state preparation while two ancilla-qubits are used for phase detection and parity detection consecutively. The left block contains a phase-checking circuit and the one in the right block is a parity-checking circuit. (b) The phase-checking circuit, contains two CNOT gates sandwiched between two Hadamard gates. (c) The parity-checking circuit contains two CNOT gates for parity detection. \label{fig:main-circuit}}
\end{figure}

We have already mentioned that in Refs. \cite{NDBSD,gupta2007general},
quantum circuit for the nondestructive discrimination of generalized
orthonormal qudit Bell states was designed by some of the present
authors (cf. Fig. 3 of \cite{NDBSD} or Fig. 4 of \cite{gupta2007general}).
As a special case of these circuits, one can easily obtain a circuit
for Bell state discrimination, as shown in Fig. \ref{fig:main-circuit}
a here and in Fig. 2 of \cite{NDBSD}. This circuit involves
4-qubits, in which the measurement on the first ancilla qubit would
reveal the phase information, whereas the second measurement would
reveal the parity information. Thus, to discriminate all four Bell
states in a single experiment, we would require a four qubit system,
allowing non-local operations between system qubits (qubits of the
Bell state) and the ancilla qubits. Apparently, this circuit should
have been implemented as it is in the 5-qubit IBM quantum computer,
but the restriction on the application of CNOT gates, restrict us
to implement this circuit as a single circuit (without causing considerable increase in gate-count and decrease in performance). Circumventing, the increase in circuit complexity (gate-count), initially, we have implemented
the phase checking circuit and the parity checking circuit separately,
as shown in Fig. \ref{fig:main-circuit} b and c, respectively. This is
consistent with the earlier NMR-based implementation of the Bell state
discrimination circuit \cite{samal2010non}, where a 3-qubit quantum
computer was used and naturally, parity checking part and the phase
checking part was performed via 2 independent experiments. In fact,
in the NMR-based implementation of the non-destructive discrimination
of Bell states, a $^{13}CHFBr_{2}$ molecule was used to perform
the quantum computing, as the number of independent Larmor frequency
of that was 3, the quantum computer was a 3-qubit one. Specifically,
Samal et al., used three nuclear spins, namely $^{1}H,\,^{13}C$ and
$^{19}F$ of $^{13}CHFBr_{2}$ which mimics a three-qubit system \cite{samal2010non}.
In their experiment, they used $^{13}C$ as ancilla qubit and rest
as system qubits. The availability of single ancilla qubit prohibited
non-destructive discrimination of the Bell states in one shot. Due
to the trade-off between the available quantum resources and number
of experiments, they used two experiments to realize the complete
protocol, one for obtaining parity information and the other one for
phase information. Circuit shown in Fig. \ref{fig:main-circuit} c is actually
used for parity checking and it is easy to observe that for Bell states
having even parity (i.e., for $|\psi^{\pm}\rangle=\frac{|00\rangle\pm|11\rangle}{\sqrt{2}}$),
the measurement on the ancilla qubit would yield $|0\rangle$ and
for the other two Bell states (i.e., for odd parity states $|\phi^{\pm}\rangle=\frac{|01\rangle\pm|10\rangle}{\sqrt{2}}$
) it would yield $|1\rangle.$ In a similar fashion, the quantum circuit
shown in Fig. \ref{fig:main-circuit} b would determine the relative phase
of the Bell state as the measurement on ancilla in Fig.
\ref{fig:main-circuit} b would yield $|0\rangle$ for $+$ states, i.e., for $|\psi^{+}\rangle$
and $|\phi^{+}\rangle$ and it would yield $|1\rangle$ for $-$states
$|\psi^{-}\rangle$ and $|\phi^{-}\rangle$ (see 3rd and 4th column
of Table \ref{tab:one} for more detail). 

\begin{figure}
\begin{centering}
\includegraphics[width=8cm]{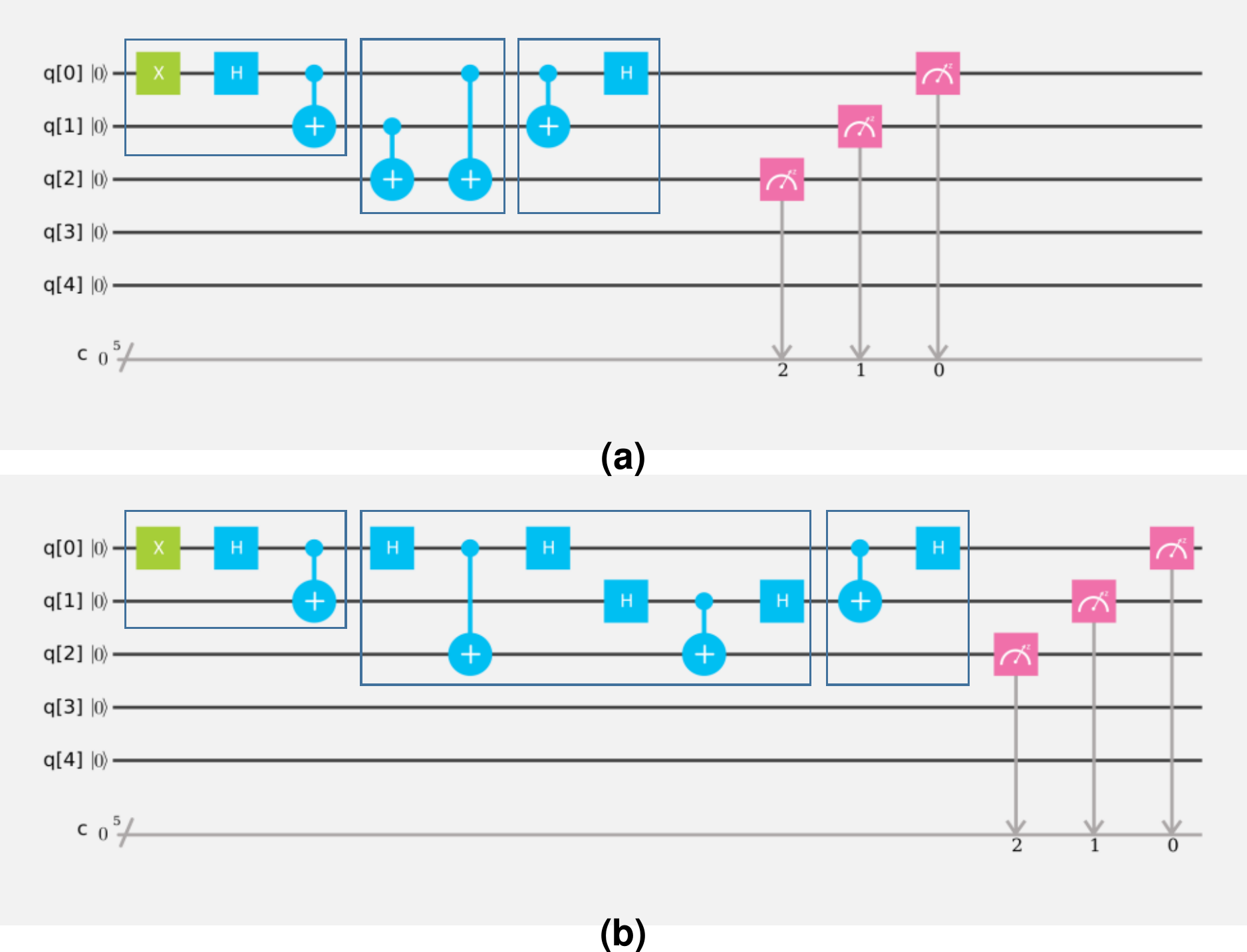}
\par\end{centering}
\caption{The counter part of parity and phase checking circuits which are experimentally implementable using five-qubit IBM quantum computer. Qubits q$\left[0\right]$, q$\left[1\right]$ are system-qubits, and $q\left[2\right]$ mimics ancilla-qubits. (a) The parity checking circuit, the left block prepares desired Bell-states (the shown configuration is for preparing $\ket{\psi^{-}}$). The middle block is a parity checking circuit which involves two CNOT gates realizable in IBM QC. The right block applies reverse EPR circuit to measure system-qubit state in computational basis. (b) The phase checking circuit, the middle block implements phase checking circuit in IBM. \label{fig:circuit-in-IBM}}

\end{figure}

\begin{figure}
\begin{centering}
\includegraphics[width=8cm]{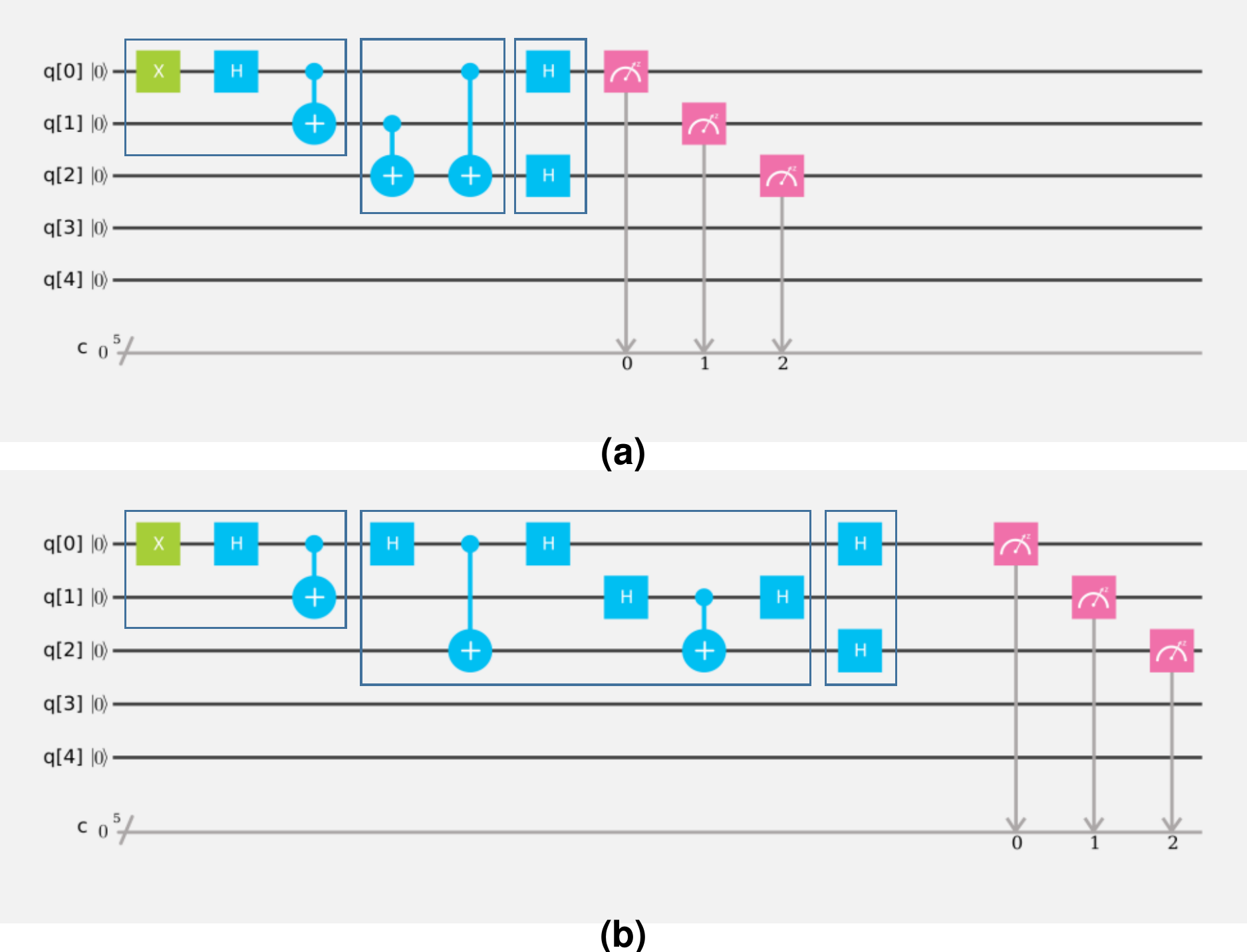}
\par\end{centering}
\caption{The counter part of (a) parity and (b) phase checking circuits which are experimentally implementable in five-qubit IBM quantum computer followed by tomography block. Tomography block involves measurement in different basis, shown configuration implements measurement in X- basis.  \label{fig:state-tomography}}
\end{figure}

We have already mentioned that the limitations of the available quantum
resources restricted Samal et al., \cite{samal2010non} to realize
the circuits shown in Fig. \ref{fig:main-circuit} b and c instead of the
whole circuit shown in Fig. \ref{fig:main-circuit} a in their liquid NMR-based
work. The same situation prevailed in the subsequent work of the same
group \cite{Anil-Kumar-paper2}. In contrast, the IBM quantum computer
is a 5-qubit one, which is SQUID based, and does not face scalability
issues encountered by NMR-based proposals. However, in the particular
case of 5-qubit IBM quantum computer, coupling is not present between
all the qubits. Absence of couplings provides restriction on the applicability
of CNOT gates. In brief, it does not allow us to perform controlled
operation between any two qubits, and consequently restricts us to
perform the nondestructive Bell state discrimination circuit either using
two independent experiments as was done in \cite{samal2010non}
(cf. Fig. 1 of \cite{samal2010non}) or using a circuit having considerably high gate-count (implementation of such a circuit will be described in the next section). Interestingly, IBM quantum
experience does not even allow us to implement the circuits shown
in Fig. \ref{fig:main-circuit} b in its actual form. We need to make some
modifications to obtain equivalent circuits. In Fig. \ref{fig:circuit-in-IBM}
a and b, we have shown the actual circuits prepared in IBM quantum
computer for parity checking and phase information checking. In both
the circuits left-most box contains an EPR circuit used for preparation
of the Bell state $|\psi^{-}\rangle$(similarly other states were
prepared and measured). The second box from the left in Fig. \ref{fig:circuit-in-IBM}
a and b provide equivalent circuits for the circuits given in Fig.
\ref{fig:circuit-in-IBM} c and b, respectively, and one can easily observe that the circuit
for obtaining the phase information required decomposition for successful
implementation in IBM quantum computer. In third box, a reverse EPR
circuit is inserted to establish that the measurement on ancilla does
not destroy the Bell state. To obtain further information about the
output states of a given circuit, quantum state tomography is done.
In Fig. \ref{fig:state-tomography}, we have shown a circuit that can be used
to perform quantum state tomography and thus to yield each
element of the density matrix of the output state before measurement
is performed. Specifically, in
Fig. \ref{fig:state-tomography}, third block from the left, two hadamard gates are applied.
This is operated to perform state tomography. To be precise, application
of a hadamard gate transforms the measurement basis from computational basis
 $\{|0\rangle,|1\rangle\}$ to diagonal basis $\{|+\rangle,|-\rangle\}$ and thus
  yields $\langle X \rangle X$ element of the single qubit density matrix. Here it would
be apt to briefly describe the method adopted here for performing
state tomography and measuring fidelity with an example. Theoretically
obtained density matrix of $|\psi^{+}\rangle|0\rangle$ is,

\begin{equation}
\begin{array}{lcc}
\rho^{T} & = & |\psi^{+}0\rangle\langle\psi^{+}0|,\end{array}\label{eq:3}
\end{equation}
where superscript $T$ indicates a theoretical (ideal) density matrix.
To check how nicely this state is prepared in experiment, we need
to reconstruct the density matrix of the output state using quantum
state tomography by following the method adopted in Refs. \cite{chuang1998bulk,
james2001measurement, hebenstreit2017compressed, alsina2016experimental,
rundle2016quantum, filipp2009two, shukla2013ancilla}. Characterization of a 3-qubit experimental density
matrix requires extraction of information from the experiments and then using that information to reconstruct experimental density matrix.
In the Pauli basis an experimental density matrix can be written as $\rho^{E}=\frac{1}{8}\underset{i,j,k}{\sum}c_{ijk}\sigma_{i}\otimes\sigma_{j}\otimes\sigma_{k},$
where $c_{ijk}=\langle\sigma_{i}\otimes\sigma_{j}\otimes\sigma_{k}\rangle$
and $\sigma_{ijk}=I,\,X,\,Y,\,Z$, and superscript $E$ is used to
indicate experimental density matrix. Reconstruction of this $8\times8$ density matrix of
3-qubit state requires knowledge of 63 unknown real parameters. The evaluation of co-efficient $c_{ijk}$
requires $\langle\sigma_{i}\rangle$ where $\sigma_{i}=I,\,X,\,Y,\,Z$ for each qubit.
 Since $\langle I \rangle$ can be obtain by the experiment done on $Z$ basis, 
instead of four we need only three measurement (measurement in $X$, $Y$, and $Z$ basis) on each qubit, thus
requiring total 27 measurements to tomograph each density matrix. Since in IBM, the only available basis for performing the measurements is $Z$ basis. To realize a 
measurement in $X$ and $Y$ basis we need to apply $H$ and
$S^{\dagger}H$ gate respectively prior to the $Z$ basis measurement and run each experiment 8192 times. 
Subsequently, to reconstruct the 3-qubit state $|\psi^{+}\rangle|0\rangle$
and to check how well the state is prepared, we have to perform 27 experiments, each of which would run 8192 times. At a later stage of the investigation, following the
same strategy, we would obtain the density matrices of the retained state
 after measuring the ancilla qubits to discriminate the
Bell states

Once $\rho^{E}$ is obtained through quantum state tomography, the
same may be used to obtain fidelity and thus a quantitative feeling
about the accuracy of the experimental implementation can be obtained.
Here it would be apt to mention that the fidelity is obtained using
the formula $Tr\sqrt{\sqrt{\rho_{1}}.\rho_{2}.\sqrt{\rho_{1}}}$ \cite{pathak2013elements}
where, $\rho_{1}$ is the density matrix of the desired state
and $\rho_{2}$ is that of the state actually prepared in the experiment.
For example, when the desired state was $|\psi^{+}0\rangle$ then
$\rho_{1}=|\psi^{+}0\rangle\langle\psi^{+}0|,$ and $\rho_{2}$ would
be obtained experimentally by performing state tomography in a manner
described above.

\begin{table*}
\begin{centering}
\begin{tabular}{|>{\centering}p{2.5cm}|>{\centering}p{2.5cm}|>{\centering}p{2cm}|>{\centering}p{2cm}|>{\centering}p{2.5cm}|>{\centering}p{2.5cm}|}
\hline 
Bell state to be discriminated/identified by nondestructive measurement & Bell-state-ancilla composite state

(considered as 3-qubit state as implemented in the experiment performed
here) & Outcome of measurement on ancilla qubit used for parity checking & Outcome of measurement on ancilla qubit used for revealing phase information & Outcome of 3-qubit measurement of parity checking circuit when the
output Bell state is measured after passing through a reverse EPR
circuit (cf. Fig. \ref{fig:all qubits measured} a-d) & Outcome of 3-qubit measurement of relative phase checking circuit
when the output Bell state is measured after passing through a reverse
EPR circuit (cf. Fig. \ref{fig:all qubits measured} e-h)\tabularnewline
\hline 
$|\psi^{+}\rangle=\frac{|00\rangle+|11\rangle}{\sqrt{2}}$ & $|\psi^{+}\rangle|0\rangle$ & $|0\rangle$ & $|0\rangle$ & $|000\rangle$ & $|000\rangle$\tabularnewline
\hline 
$|\psi^{-}\rangle=\frac{|00\rangle-|11\rangle}{\sqrt{2}}$ & $|\psi^{-}\rangle|0\rangle$ & $|0\rangle$ & $|1\rangle$ & $|100\rangle$ & $|101\rangle$\tabularnewline
\hline 
$|\phi^{+}\rangle=\frac{|01\rangle+|10\rangle}{\sqrt{2}}$ & $|\phi^{+}\rangle|0\rangle$ & $|1\rangle$ & $|0\rangle$ & $|011\rangle$ & $|010\rangle$\tabularnewline
\hline 
$|\phi^{-}\rangle=\frac{|01\rangle-|10\rangle}{\sqrt{2}}$ & $|\phi^{-}\rangle|0\rangle$ & $|1\rangle$ & $|1\rangle$ & $|111\rangle$ & $|111\rangle$\tabularnewline
\hline 
\end{tabular}
\par\end{centering}
\caption{Table shows expected outcomes after parity checking circuit and phase checking circuit for Bell-state-ancilla composite system for all Bell states. Expected outputs of ancilla qubits in the same cases are also given. In particular, Column 1 shows Bell states to be examined, Column 2 shows states of the Bell-state-ancilla composite system, Columns 3 and 4 show outcomes of measurements on ancilla  for two cases, and Columna 5 and 6 show the outcomes of measuring composite states in the computational basis (from left to right). \label{tab:one}}
\end{table*}

\section{Results\label{sec:Results}}

To begin with we prepare 4 Bell states (using EPR circuit, i.e., a
hadamard followed by CNOT) and an ancilla in state $|0\rangle$. It
is well known that an EPR circuit transforms input states $|00\rangle,|01\rangle,|10\rangle,$
$|11\rangle$ into $|\psi^{+}\rangle=\frac{|00\rangle+|11\rangle}{\sqrt{2}}$,$\,|\phi^{+}\rangle=\frac{|01\rangle+|10\rangle}{\sqrt{2}},\,|\psi^{-}\rangle=\frac{|00\rangle-|11\rangle}{\sqrt{2}}$
and $|\phi^{-}\rangle=\frac{|01\rangle-|10\rangle}{\sqrt{2}}$, respectively.
Default initial state in IBM quantum experience is $|0\rangle$ for
each qubit line. However, the input states required by an EPR circuit
to generate different Bell states can be prepared by placing ${\rm NOT}$
gate(s) in appropriate positions before the EPR circuit (see how $|\psi^{-}\rangle$
is prepared in left most block of Fig. \ref{fig:circuit-in-IBM} ). This
is how Bell states are prepared here. To check the accuracy of the
states prepared in the experiments, quantum state tomography of 
the experimentally obtained density matrices are performed. For this purpose, we have followed
the method described in the previous section. Density matrix for the
experimentally obtained state corresponding to a particular case (for
the expected state $|\psi^{+}0\rangle,$ i.e., for $\text{\ensuremath{\rho}}^{T}=\rho_{1}=|\psi^{+}0\rangle\langle\psi^{+}0|),$
is obtained as 
\begin{equation}
\rho_{|\psi^{+}0\rangle}^{E}={\rm Re}\left[\rho_{|\psi^{+}0\rangle}^{E}\right]+i{\rm \,Im}\left[\rho_{|\psi^{+}0\rangle}^{E}\right],\label{eq:rho1}
\end{equation}
where a subscript is added to uniquely connect the experimental density
matrix with the corresponding ideal state and
\begin{widetext}
\begin{equation}
\begin{array}{lcc}
{\rm Re}\left[\rho_{|\psi^{+}0\rangle}^{E}\right] & = & \mbox{\mbox{\ensuremath{\left(\begin{array}{cccccccc}
 0.4411  &  0.003  &  0.011  &  -0.0102  &  0.006  &  -0.005  &  0.3657  &  0.0065\\
 0.003  &  0.0021  &  0.0137  &  -0.0005  &  0.0045  &  0.0005  &  0.011  &  0.0015\\
 0.011  &  0.0137  &  0.0741  &  0.003  &  0.0047  &  -0.004  &  0.006  &  0.0017\\
 -0.0102  &  -0.0005  &  0.003  &  0.0001  &  -0.0015  &  -0.0005  &  -0.0012  & 0\\
 0.006  &  0.0045  &  0.0047  &  -0.0015  &  0.0731  &  0.001  &  0.0035  &  -0.0037\\
 -0.005  &  0.0005  &  -0.004  &  -0.0005  &  0.001  &  0.0001  &  0.0112  &  -0.0005\\
 0.3657  &  0.011  &  0.006  &  -0.0012  &  0.0035  &  0.0112  &  0.4081  &  0.01\\
 0.0065  &  0.0015  &  0.0017  &  0  &  -0.0037  &  -0.0005  &  0.01  &  0.0011 
\end{array}\right)}}}\end{array}\label{eq:rhoreal}
\end{equation}
 and 
\begin{equation}
\begin{array}{lcc}
{\rm Im}\left[\rho_{|\psi^{+}0\rangle}^{E}\right] & = & \left(\begin{array}{cccccccc}
0 & -0.018 & -0.0235 & -0.0267 & -0.0335 & -0.003 & -0.0302 & -0.0176\\
0.018 & 0 & -0.0027 & 0 & -0.0015 & 0 & 0.0213 & 0\\
0.0235 & 0.0027 & 0 & 0.0175 & 0.0302 & -0.0023 & -0.01 & 0.006\\
0.0267 & 0 & -0.0175 & 0 & 0.0036 & -0.0005 & 0.0032 & 0\\
0.0335 & 0.0015 & -0.0302 & -0.0036 & 0 & -0.0045 & -0.0025 & -0.0227\\
0.003 & 0 & 0.0023 & 0.0005 & 0.0045 & 0 & -0.0007 & 0\\
0.0302 & -0.0213 & 0.01 & -0.0032 & 0.0025 & 0.0007 & 0 & -0.0025\\
0.0176 & 0 & -0.0062 & 0 & 0.0227 & 0 & 0.0025 & 0
\end{array}.\right)\end{array}\label{eq:rho-imaginary}
\end{equation}
\end{widetext}

\begin{figure}
\begin{centering}
\includegraphics[width=8cm]{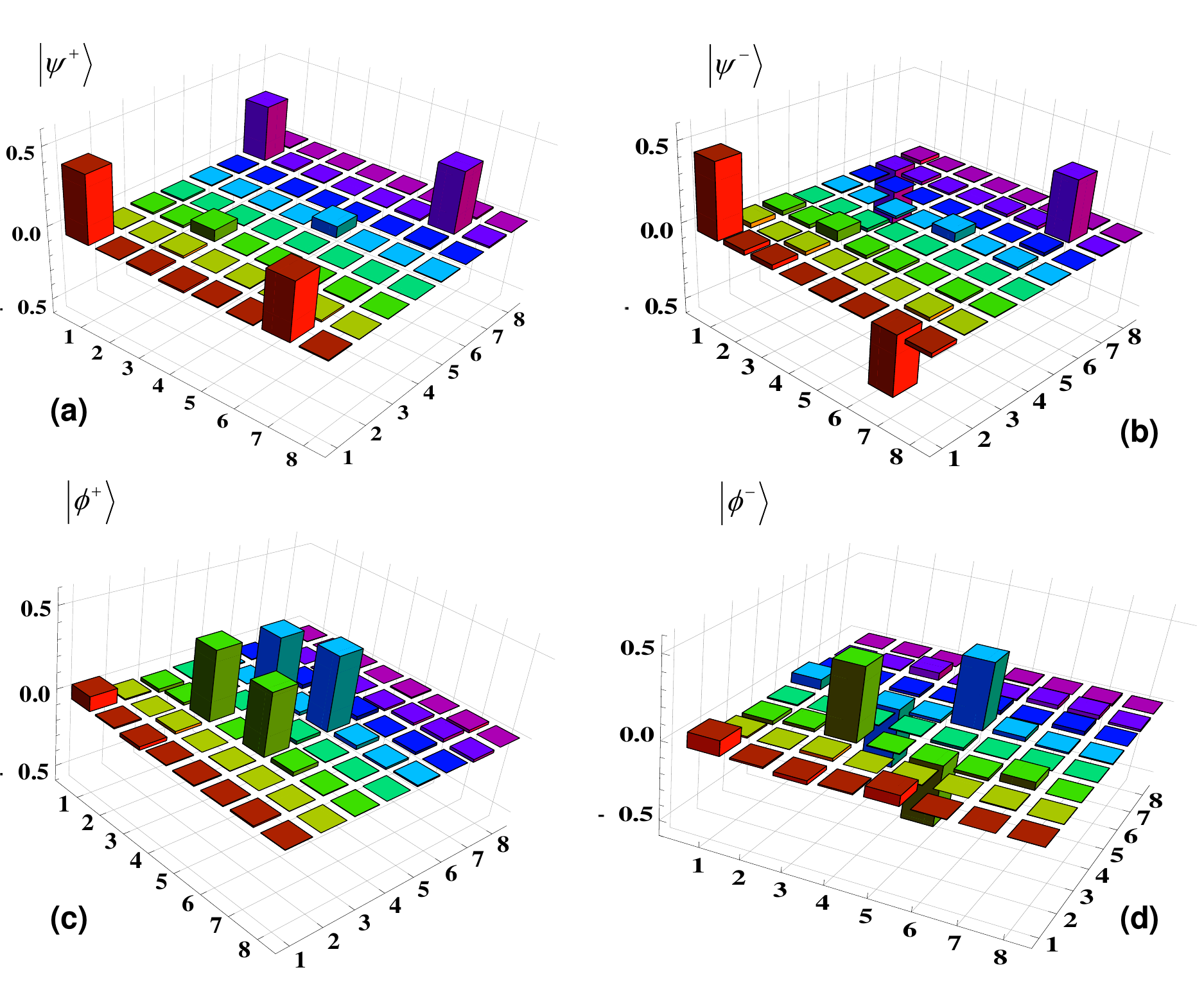}
\par\end{centering}
\caption{ Reconstructed Bell-states on Bell-state-ancilla composite
system corresponding to ideal states (a) $\ket{\psi^{+}0}$, (b) $\ket{\psi^{-}0}$, (c) $\ket{\phi^{+}0}$, (d) and $\ket{\phi^{-}0}$. 
In each plot of the figure, the states $|000\rangle$,\,$|001\rangle$,\,$|010\rangle$,\,$|011\rangle$,\,$|100\rangle$,\,$|101\rangle$,\,$|110\rangle$\,and\,$|111\rangle$ are lebelled as 1-8 consecutively in $X$ and $Y$ axis.
\label{fig:bell-reconstruct}}
\end{figure}

Real part of this density matrix is illustrated in Fig. \ref{fig:bell-reconstruct}
a. Figure \ref{fig:bell-reconstruct} also illustrates the real part
of density matrices of the experimentally prepared Bell-state-ancilla
composite in the other three cases. Corresponding density matrices are
provided in Appendix 1 (see Eq. (\ref{eq:new1})-(\ref{eq:new6})). Figure \ref{fig:bell-reconstruct} clearly
shows that the Bell states are prepared with reasonable amount of
accuracy, but does not provide any quantitative measure of accuracy.
So we have calculated the 'average absolute deviation' $\langle\Delta x\rangle$
and 'maximum absolute deviation' $\Delta x_{max}$ of the experimental density
matrix from the theoretical one by using this formulae 

\begin{equation}
\begin{array}{lcc}
\langle\Delta x\rangle & = & \frac{1}{N^{2}}\stackrel[i,j=1]{N}{\sum|x_{i,j}^{T}}-x_{i,j}^{E}|,\\
\Delta x_{max} & = & Max|x_{i,j}^{T}-x_{i,j}^{E}|,\\
\forall i,\,j & \in & \left\{ 1,\,N\right\} .
\end{array}\label{eq:5}
\end{equation}
where, $x_{ij}^{T}$ and $x_{ij}^{E}$ are the theoretical and experimental
elements. Putting the values in this equation from
eq3 and eq4 we find the 'average absolute deviation' and the 'maximum
absolute deviation' of the Bell state $|\psi^{+}\rangle|0\rangle$
which is $1.8\%$ and $13.7\%$. 

Similarly, we have also calculated 'average absolute deviation' and
'maximum absolute deviation' for other Bell states too, these values are 
$|\psi^{-}\rangle|0\rangle$is $1.8\%$ and $12.5\%$,
$|\phi^{+}\rangle$ is $1.8\%$ and $11.9\%$ and $|\phi^{-}\rangle$
is $2\%$ and $11.8\%$, respectively. It may be noted that in NMR-based
experiment \cite{samal2010non}, the values for 'average absolute
deviation' and 'maximum absolute deviation' is reported as $\sim1\%$
and $\sim4\%$ respectively. This probably indicates, that Bell states
were prepared in NMR-based experiment, with higher accuracy, but the
measure used there was not universal. So we also compute fidelity
of the reconstructed state with respect to the desired state and obtained
following values of fidelity for various Bell-state-ancilla composite
system: $F_{|\psi^{+}0\rangle}=0.8890,\,F_{|\psi^{-}0\rangle}=0.8994,\,F_{|\phi^{+}0\rangle}=0.9091,$
and $F_{|\phi^{-}0\rangle}=0.9060,$ where $F_{|i\rangle}$ denotes
the fidelity at which the desired quantum state $|i\rangle$ is prepared
in the IBM quantum computer. The computed values of $F_{|i\rangle}$
clearly show that the Bell-state-ancilla composite system has been
prepared with considerable accuracy in the present experiment.

\begin{figure}
\begin{centering}
\includegraphics[width=8cm]{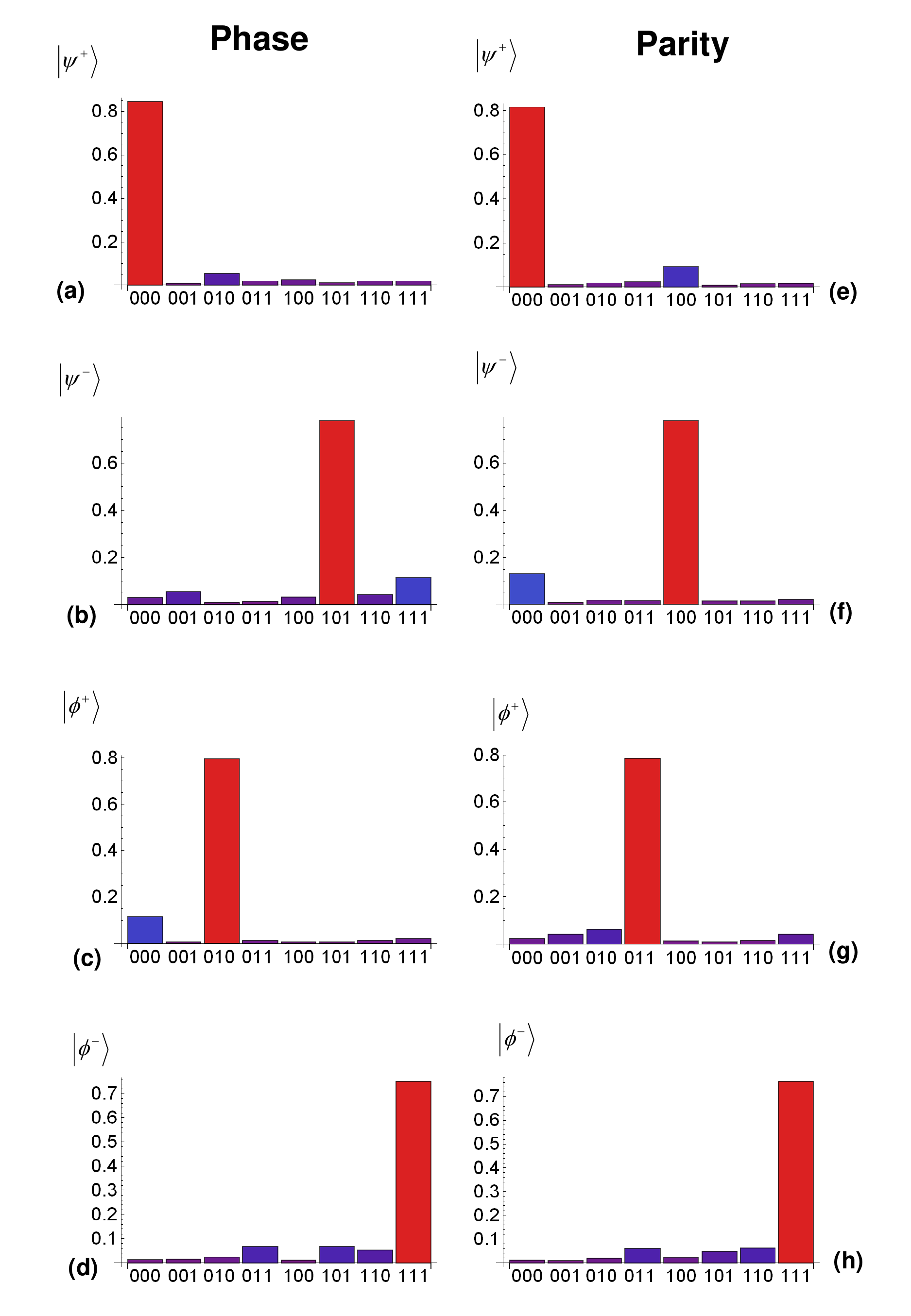}
\par\end{centering}
\caption{Experimental results after implementation of circuits in Fig. \ref{fig:circuit-in-IBM}. (a-d) Measurement results after implementing phase checking circuit. (e-h) Measurement results after implementing phase checking circuit. \label{fig:all qubits measured}}
\end{figure}

After verifying that the Bell-state-ancilla composite system are prepared
successfully, we perform measurements on the ancilla to perform nondestructive
discrimination of the Bell state. After, the measurement of ancilla,
the Bell state is expected to remain unchanged, to test that a
reverse EPR circuit is applied to the system qubits, and subsequently
the system qubits are measured in computational basis. The reverse
EPR circuit actually transforms a Bell measurement into a measurement
in the computational basis. Outcomes of these measurements are shown
in Fig. \ref{fig:all qubits measured}, within the experimental error, these results are consistent with the expected theoretical results shown in Column
4 and 5 of Table \ref{tab:one}. Thus, nondestructive discrimination is successfully performed in IBM quantum computer.

Figure \ref{fig:all qubits measured} definitely indicates a successful
implementation of nondestructive discrimination of Bell states. However,
it does not reveal the whole picture. To obtain the full picture,
we perform quantum state tomography after implementation of parity checking circuit
and phase information checking circuits. 
In the following, we provide experimental density matrices ${\normalcolor {\left[\rho^{E}{}_{|\psi^{+}0\rangle}\right]}}$ for both cases (phase and parity) corresponding to the ideal state $|\psi^{+}0\rangle$,
obtained through quantum sate tomography.

\begin{widetext}
\begin{equation}
\begin{array}{lcc}
{\normalcolor {\rm Re}\left[\rho^{E}{}_{|\psi^{+}0\rangle}\right]_{parity}} & = & {\color{black}\left(\begin{array}{cccccccc}
0.4621 & 0.0845 & 0.0105 & 0.0187 & 0.0155 & 0.0167 & 0.3352 & 0.0148\\
0.0845 & 0.0051 & 0.0537 & -0.005 & 0.0077 & 0.001 & 0.0793 & -0.0017\\
0.0105 & 0.0537 & 0.0361 & -0.0315 & -0.0047 & 0.0011 & 0.009 & 0.009\\
0.0187 & -0.005 & -0.0315 & 0.0291 & -0.0018 & 0.0022 & 0.024 & 0.002\\
0.0155 & 0.0077 & -0.0047 & -0.0018 & 0.0461 & 0.012 & 0.0085 & 0.006\\
0.0167 & 0.001 & 0.0011 & 0.0022 & 0.012 & 0.0101 & 0.0565 & -0.003\\
0.3352 & 0.0793 & 0.009 & 0.024 & 0.0085 & 0.0565 & 0.3991 & 0.0115\\
0.0148 & -0.0017 & 0.009 & 0.002 & 0.006 & -0.003 & 0.0115 & 0.0121
\end{array}\right)},\end{array}\label{eq:rhoparity1}
\end{equation}

\begin{equation}
\begin{array}{lcc}
{\rm Im}\left[\rho^{E}{}_{|\psi^{+}0\rangle}\right]_{parity} & = & {\color{black}\left(\begin{array}{cccccccc}
0 & -0.0545 & -0.0065 & -0.0575 & -0.0105 & -0.0107 & -0.148 & -0.0837\\
0.0545 & 0 & 0.007 & 0.006 & -0.0047 & -0.001 & 0.0055 & -0.0002\\
0.0065 & -0.007 & 0 & 0.0465 & -0.0105 & -0.0057 & -0.0085 & 0.0005\\
0.0575 & -0.006 & -0.0465 & 0 & 0 & 0.001 & -0.0085 & -0.004\\
0.0105 & 0.0047 & 0.0105 & 0 & 0 & -0.011 & -0.0115 & -0.0387\\
0.0107 & 0.001 & 0.0057 & -0.0017 & 0.011 & 0 & -0.0027 & 0.0045\\
0.148 & -0.0055 & 0.0085 & 0.0085 & 0.0115 & 0.0027 & 0 & -0.031\\
0.0837 & 0.0002 & -0.0005 & 0.004 & 0.0387 & -0.0045 & 0.031 & 0
\end{array}\right)},\end{array}\label{eq:rhoparity2}
\end{equation}

\begin{equation}
\begin{array}{lcc}
{\normalcolor {\rm Re}\left[\rho^{E}{}_{|\psi^{+}0\rangle}\right]_{phase}} & {\normalcolor =} & {\normalcolor \left(\begin{array}{cccccccc}
0.4098 & 0.058 & 0.038 & 0.0437 & 0.0385 & 0.028 & 0.36625 & 0.0353\\
0.058 & 0.0148 & 0.0522 & -0.007 & 0.0055 & 0 & 0.0523 & 0.001\\
0.038 & 0.0522 & 0.0648 & -0.034 & 0.0247 & 0.0188 & 0.03 & -0.0062\\
0.0437 & -0.007 & -0.034 & 0.0258 & 0.0108 & -0.0035 & 0.0382 & -0.004\\
0.0385 & 0.0055 & 0.0247 & 0.0108 & 0.0728 & 0.0145 & 0.037 & -0.0042\\
0.028 & 0 & 0.0188 & -0.0035 & 0.0145 & 0.0098 & 0.0677 & -0.005\\
0.3662 & 0.0523 & 0.03 & 0.0382 & 0.037 & 0.0677 & 0.3738 & 0.012\\
0.0353 & 0.001 & -0.0062 & -0.004 & -0.0042 & -0.005 & 0.012 & 0.0278
\end{array}\right),}\end{array}\label{eq:rhophase1}
\end{equation}

\begin{equation}
\begin{array}{lcc}
{\normalcolor {\rm Im}\left[\rho^{E}{}_{|\psi^{+}0\rangle}\right]_{phase}} & {\normalcolor =} & {\normalcolor {\normalcolor {\normalcolor \left(\begin{array}{cccccccc}
0 & -0.0405 & 0.0395 & -0.0342 & 0.019 & 0.0055 & -0.021 & -0.0805\\
0.0405 & 0 & 0.0157 & 0.0055 & 0.0085 & 0 & 0.0222 & 0.0012\\
-0.0395 & -0.0157 & 0 & 0.043 & -0.0015 & -0.0142 & -0.0545 & 0.0005\\
0.0342 & -0.0055 & -0.043 & 0 & 0.0135 & 0.0027 & -0.02 & -0.01\\
-0.019 & -0.0085 & 0.0015 & -0.0135 & 0 & -0.0085 & -0.0585 & -0.0545\\
-0.0055 & 0 & 0.0142 & -0.0027 & 0.0085 & 0 & -0.0165 & 0.0065\\
0.021 & -0.0222 & 0.0545 & 0.02 & 0.0585 & 0.0165 & 0 & -0.0435\\
0.0805 & -0.0012 & -0.0005 & 0.01 & 0.0545 & -0.0065 & 0.0435 & 0
\end{array}\right)}}}\end{array}\label{eq:rhophase2}
\end{equation}
\end{widetext}

\begin{figure}
\begin{centering}
\includegraphics[width=8cm]{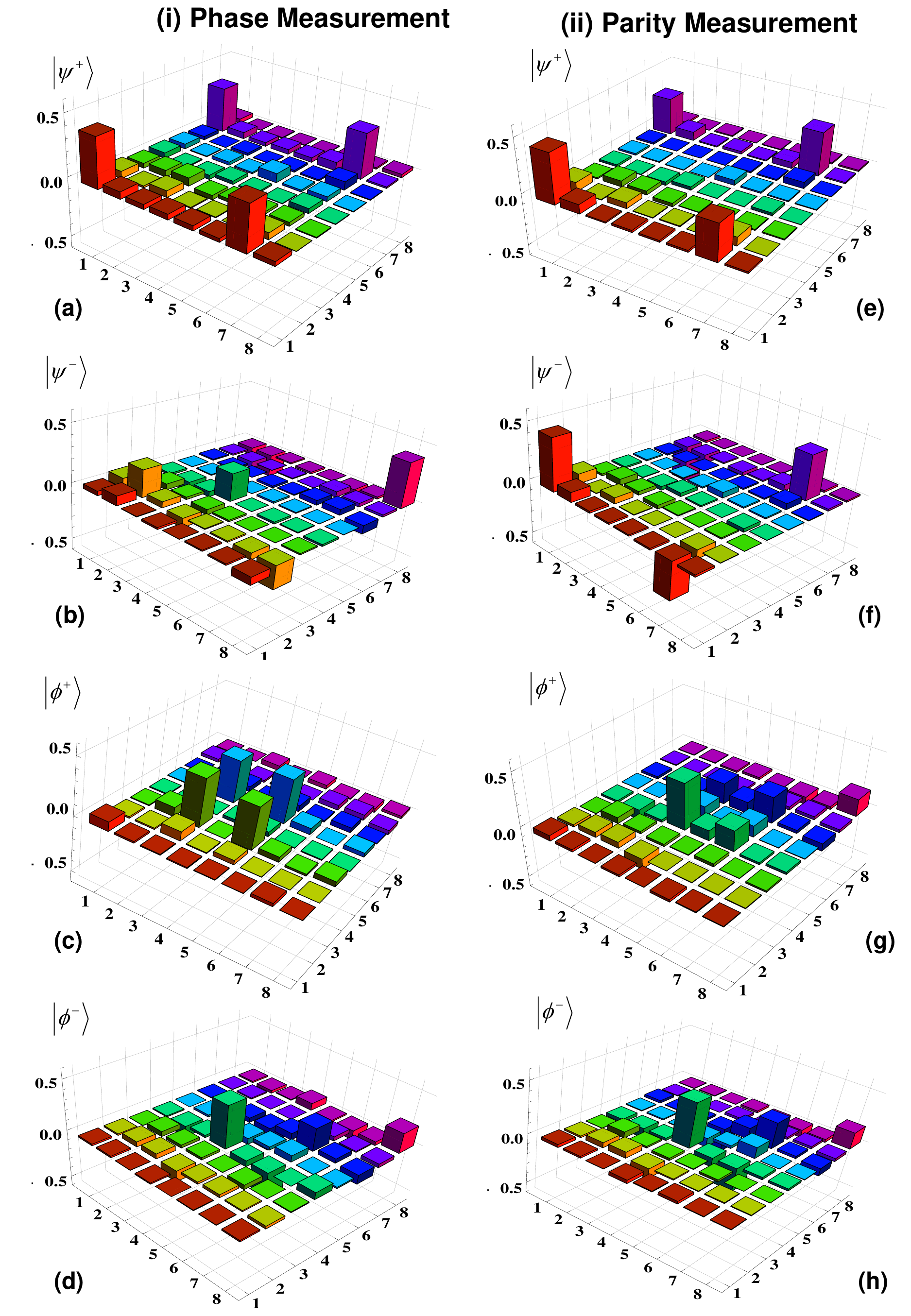}
\par\end{centering}
\caption{Reconstructed density matrices of various states of ancilla Bell-state combined system obtained after implementing circuit in Fig. \ref{fig:state-tomography}. The first column (a-d) contains density matrices  
corresponding to ideal states $\ket{\psi^{+}0}$, $\ket{\psi^{-}1}$, $\ket{\phi^{+}0}$, and $\ket{\phi^{-}1}$ respectively. The density matrices in the second column corresponds to ideal states (e-h) $\ket{\psi^{+}0}$, $\ket{\psi^{-}0}$, $\ket{\phi^{+}1}$, and $\ket{\phi^{-}1}$ respectively. \label{fig:Final-fig}}
\end{figure}

\begin{figure*}
\begin{centering}
\includegraphics[width=10cm]{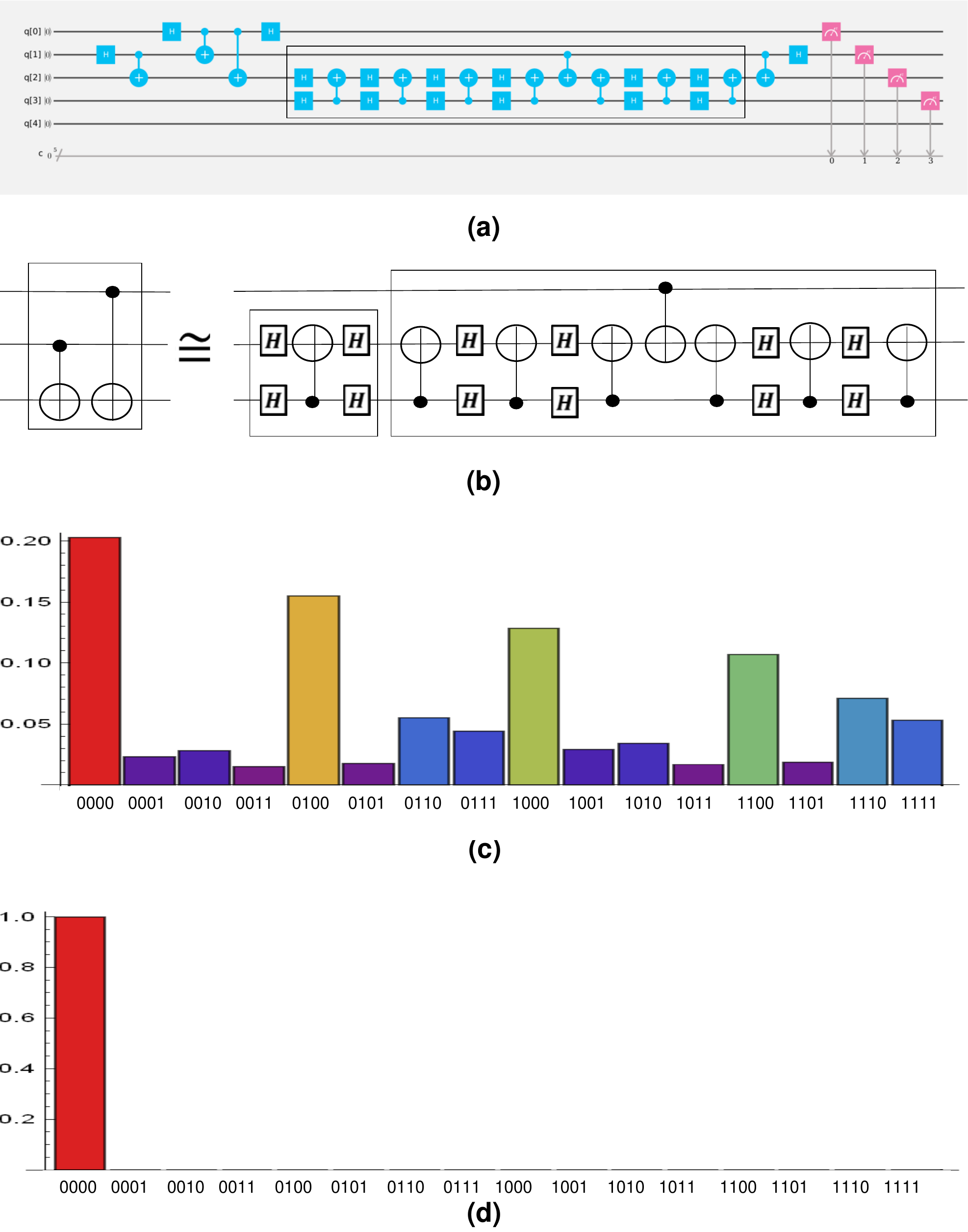}
\par\end{centering}
\caption{(a) Actual implementation of the 4-qubit quantum circuit shown in Fig. \ref{fig:main-circuit} a  using IBM quantum computer (for the Bell state $\ket{\psi^{+}}$. (b) The  implemented circuit utilizes this circuit identity to circumvent the constraints that are present in IBM quantum computer. (c) Probability of various measurement outcomes  obtained after 8192 runs of the circuit in actual quantum computer. (d) Simulated outcome after running circuit on IBM simulator. This outcome agrees exactly with the theoretically expected value, but the experimental outcome shown in 
(c) considerably deviates from it. \label{fig:fig4}}
\end{figure*}

The subscripts ``parity'' and ``phase'' denotes the experiment
for which the experimental density matrix is obtained via quantum
state tomography. Corresponding density matrices for the other Bell-state-Ancilla composites are reported in Appendix 1 (Eq. (\ref{eq:new7})-(\ref{eq:new18})). Real part of the density matrices
obtained through the parity checking and phase information checking
circuits are shown in Fig. \ref{fig:Final-fig}. The results illustrated
through these plots clearly show that the Bell state discrimination
has been realized appropriately. Further, the obtained density matrices
allows us to quantitatively establish this fact through the computation
of fidelity, and analogy of Fig. \ref{fig:Final-fig} with Fig. 6
of Ref. \cite{samal2010non} allows us to compare the NMR-based
results with the SQUID-based results. However, the  nonavailability of the
exact density matrices for the NMR-based results, restricts us from
a quantitative comparison. The obtained fidelities for the realization of phase and parity information checking circuits  are given below. 
The corresponding cases can be identified by superscript phase and parity. 
$F^{phase}_{|\psi^{+}0\rangle}$=$0.8707$, 
$F^{phase}_{|\psi^{-}1\rangle}$=$0.7114$, 
$F^{phase}_{|\phi^{+}0\rangle}$=$0.8794$, 
$F^{phase}_{|\phi^{-}1\rangle}$=$0.7493$,
$F^{parity}_{|\psi^{+}0\rangle}$=$0.8751$, 
$F^{parity}_{|\psi^{-}0\rangle}$=$0.8751$, 
$F^{parity}_{|\phi^{+}1\rangle}$=$0.7224$, and 
$F^{parity}_{|\phi^{-}1\rangle}$=$0.7576$, here the ideal state is given in subscript and 
superscript "phase" and "parity" corresponds to phase discrimination and parity discrimination
realized by phase checking circuit in Fig. \ref{fig:state-tomography} a and parity checking circuit
in Fig. \ref{fig:state-tomography} b. Obtained fidelities are reasonably good, but to make a SQUID-based scalabale quantum computer, we have to considerably improve the quality of the quantum gates. Specifically, we can see that the fildelities of the constructed Bell states were much higher than the fidelities obtained after phase information or parity information of the given Bell state is obtained through the distributed measurement. Clearly, increase in circuit complexity has resulted in the reduction of fidelity. To illustrate this point, we would now report implementation of circuit shown in Fig. \ref{fig:main-circuit} a, i.e., implementation
of a 4-qubit circuit for non-destructive discrimination of Bell state, where phase information and parity information will be revealed in a single experiment. A 4-qubit quantum circuit corresponding to the circuit shown in Fig. \ref{fig:main-circuit} a with initial Bell state $|\psi^{+}\rangle$ is implemented using IBM quantum computer and the same is shown in Fig. \ref{fig:fig4} a, where we have used a circuit theorem shown in Fig. \ref{fig:fig4} b. Although,  the circuit and the corresponding results shown here are for $\ket{\psi^{+}}$, but we have performed experiments for all possible Bell states and have obtained similar results (which are not illustrated here). Due to the restrictions provided by the IBM computer computer, the left (right) CNOT gate shown in LHS of Fig. \ref{fig:fig4} b is implemented using the gates shown in the left (right) rectangular box shown in RHS of Fig. \ref{fig:fig4} b. In fact, the right most rectangular box actually swaps qubits 2 and 3, apply a CNOT with control at first qubit and target at the second qubit and again swaps  qubit 2 and 3. The use of the circuit identity Fig. \ref{fig:fig4} b, allows us to implement the circuit shown in Fig. \ref{fig:main-circuit} a, but it causes 10 fold increase in gate count (from 2 CNOT gates to a total of 20 gates)  for the parity checking circuit.  As a consequence of the increase in gate count, the success probability of the experiment reduces considerably, and that can be seen easily by comparing the outcome of the real experiment illustrated in Fig. \ref{fig:fig4} c with the outcome of simulation (expected state in the ideal noise-less situation) shown in Fig. \ref{fig:fig4} d.  This comparison in general and the outcome observed in Fig. \ref{fig:fig4} c, clearly illustrate that until now the technology used in IBM quantum computer is not good enough for the realization of complex quantum circuits.

\section{Conclusion\label{sec:Conclusion}}

We have already noted that nondestructive discrimination of Bell states
have wide applicability. Ranging from quantum error correction to
measurement-based quantum computation, quantum communication in a
network involving multiple parties to optimization of the quantum
communication complexity for performing measurements in distributed
quantum computing. Keeping that in mind, here, we report an experimental
realization of a scheme for nondestructive Bell state discrimination.
Due to the limitations of the available quantum resources (IBM quantum
computer being a 5-qubit quantum computer with some restriction) this
study is restricted to the discrimination of Bell states only, but
our earlier theoretical proposal is valid in general for discrimination
of generalized orthonormal qudit Bell states. We hope that the present
experiment will soon be extended to the experimental discrimination
of more complex entangled states. Further, as the work provides a
clean prescription for using IBM quantum experience to experimentally
realize quantum circuits that may form building blocks of a real quantum
computer, a similar approach may be used to realize a set of other
important circuits. Finally, the comparison with the NMR-based technology,
reveals that this SQUID-based quantum computer's performance is comparable
to that of the NMR-based quantum computer as far as the discrimination
of Bell states is concerned. As the detail of the density matrix obtained
(through quantum state tomography) in earlier works \cite{samal2010non,Anil-Kumar-paper2}
was not available, fidelity of NMR based realization earlier and the
SQUID-based realization reported here could not be compared. However,
the fidelity computed for the states prepared and retained after the
nondestructive discrimination operation is reasonably high and that
indicate the accuracy of the IBM quantum computer. Further, it is
observed that all the density matrices produced here through the state
tomography are mixed state (i.e., for all of them $Tr(\rho^{2})$<1).
This puts little light on the nature of noise present in the channel
and/or the errors introduced by the gates used. However, it can exclude
certain possibilities. For example, it excludes the possibility that
the combined effect of noise/error present in the circuit can be viewed
as bit flip and/or phase flip error as such errors would have kept
the state as pure. More on characterization of noise present in IBM
quantum experience will be discussed elsewhere.

\textbf{Acknowledgment:} AP and AS thank Defense Research $\&$ Development Organization (DRDO), India for the support provided through the project number ERIP/ER/1403163/M/01/1603.

\bibliography{ref}


\pagebreak
\begin{widetext}

\section*{Appendix1: Experimentally obtained density matrices for system-ancilla state \label{A}}
\setcounter{equation}{0}\renewcommand{\theequation}{A\arabic{equation}}
Density matrices after the reconstruction of the Bell-state-ancilla composite states $|\psi^{-}0\rangle,\,|\phi^{+}0\rangle\,$ and $\,|\phi^{-}0\rangle$
are 

\begin{equation}
\begin{array}{lcc}
{\rm Re}\left[\rho_{|\psi^{-}0\rangle}^{E}\right] & = & \left(\begin{array}{cccccccc}
0.476 & 0.029 & 0.029 & -0.0007 & -0.0065 & -0.0045 & -0.3745 & -0.0236\\
0.029 & 0.003 & 0.0222 & 0.0005 & 0.002 & 0.0 & -0.0208 & -0.0025\\
0.029 & 0.0222 & 0.066 & -0.02 & -0.015 & -0.0013 & -0.0115 & 0.0022\\
-0.0007 & 0.0005 & -0.02 & 0.001 & 0.0033 & -0.0005 & -0.0007 & 0.0\\
-0.0065 & 0.002 & -0.015 & 0.0033 & 0.058 & 0.0045 & 0.0175 & 0.0022\\
-0.0045 & 0.0 & -0.0013 & -0.0005 & 0.0045 & 0.0 & 0.0202 & 0.0\\
-0.3745 & -0.0208 & -0.0115 & -0.0007 & 0.0175 & 0.0202 & 0.393 & 0.0055\\
-0.0236 & -0.0025 & 0.0022 & 0.0 & 0.0022 & 0.0 & 0.0055 & 0.003
\end{array}\right)\end{array}\label{eq:new1}
\end{equation} 

\begin{equation}
\begin{array}{lcc}
{\rm Im}\left[\rho_{|\psi^{-}0\rangle}^{E}\right] & = & \left(\begin{array}{cccccccc}
0 & -0.0215 & -0.01 & -0.019 & -0.009 & 0 & -0.0067 & 0.0185\\
0.0215 & 0 & 0.0015 & -\text{2.16\ensuremath{\times}1\ensuremath{0^{-19}}} & -0.0005 & -0.0005 & -0.0182 & 0.001\\
0.01 & -0.0015 & 0 & 0.0225 & -0.0127 & -0.0027 & -0.012 & -0.0005\\
0.019 & 0 & -0.0225 & 0 & 0.0035 & 0 & -0.0005 & 0\\
0.009 & 0.0005 & 0.0127 & -0.0035 & 0 & -0.0025 & -0.012 & -0.0147\\
0 & 0.0005 & 0.0027 & 0 & 0.0025 & 0.0 & -0.0042 & -0.0005\\
0.0067 & 0.0182 & 0.012 & 0.0005 & 0.012 & 0.0042 & 0 & 0.001\\
-0.0185 & -0.001 & 0.0005 & 0 & 0.0147 & 0.0005 & -0.001 & 0
\end{array}\right)\end{array}\label{eq:new2}
\end{equation}

\begin{equation}
\begin{array}{lcc}
{\rm Re}\left[\rho_{|\phi^{+}0\rangle}^{E}\right] & = & \left(\begin{array}{cccccccc}
0.089 & 0.0065 & -0.0205 & 0.0035 & -0.0055 & -0.0037 & 0.008 & -0.0017\\
0.0065 & 0.0 & 0.0155 & 0.0 & -0.0017 & 0.0 & -0.0005 & 0.0\\
-0.0205 & 0.0155 & 0.429 & -0.001 & 0.3825 & 0.0242 & 0.0035 & 0.0015\\
0.0035 & 0.0 & -0.001 & 0.001 & 0.02 & 0.0025 & 0.002 & 0.0\\
-0.0055 & -0.0017 & 0.3825 & 0.02 & 0.459 & 0.024 & -0.011 & 0.0075\\
-0.0037 & 0.0 & 0.0242 & 0.0025 & 0.024 & 0.002 & 0.0145 & -0.0005\\
0.008 & -0.0005 & 0.0035 & 0.002 & -0.011 & 0.0145 & 0.02 & -0.0165\\
-0.0017 & 0.0 & 0.0015 & 0.0 & 0.0075 & -0.0005 & -0.0165 & 0.0
\end{array}\right)\end{array}\label{eq:new3}
\end{equation}

\begin{equation}
\begin{array}{lcc}
{\rm Im}\left[\rho_{|\phi^{+}0\rangle}^{E}\right] & = & \left(\begin{array}{cccccccc}
0 & -0.007 & -0.0165 & -0.0162 & -0.03 & 0.0002 & 0.0002 & 0.0025\\
0.007 & 0.0 & 0.0012 & 0 & 0.0002 & -0.0005 & -0.0027 & -0.0002\\
0.0165 & -0.0012 & 0 & -0.0095 & -0.0222 & -0.0245 & -0.017 & 0.0005\\
0.0162 & 0 & 0.0095 & 0 & 0.0192 & 0.0002 & 0.0045 & 0.0005\\
0.03 & -0.0002 & 0.0222 & -0.0192 & 0 & -0.0235 & -0.011 & -0.011\\
-0.0002 & 0.0005 & 0.0245 & -0.0002 & 0.0235 & 0 & 0.0025 & -0.0005\\
-0.0002 & 0.0027 & 0.017 & -0.0045 & 0.011 & -0.0025 & 0 & 0.0125\\
-0.0025 & 0.0002 & -0.0005 & -0.0005 & 0.011 & 0.0005 & -0.0125 & 0
\end{array}\right)\end{array}\label{eq:new4}
\end{equation}

\begin{equation}
\begin{array}{lcc}
{\rm Re}\left[\rho_{|\phi^{-}0\rangle}^{E}\right] & = & \left(\begin{array}{cccccccc}
0.092 & 0.009 & -0.023 & -0.0072 & 0.063 & 0.002 & 0 & -0.0011\\
0.009 & 0.001 & 0.0177 & 0 & 0.0035 & 0 & 0.0046 & -0.00175\\
-0.023 & 0.0177 & 0.454 & 0.0045 & -0.384 & -0.0176 & 0.047 & -0.0005\\
-0.0072 & 0 & 0.0045 & 0.005 & -0.0173 & -0.0027 & 0.001 & 0.001\\
0.063 & 0.0035 & -0.384 & -0.0173 & 0.42 & 0.0255 & -0.009 & -0.0052\\
0.002 & 0 & -0.0176 & -0.0027 & 0.0255 & 0.003 & 0.0252 & 0\\
0 & 0.0046 & 0.047 & 0.001 & -0.009 & 0.02525 & 0.025 & -0.013\\
-0.0011 & -0.0017 & -0.0005 & 0.001 & -0.0052 & 0 & -0.013 & 0
\end{array}\right)\end{array}\label{eq:new5}
\end{equation}

\begin{equation}
\begin{array}{lcc}
{\rm Im}\left[\rho_{|\phi^{-}0\rangle}^{E}\right] & = & \left(\begin{array}{cccccccc}
0 & -0.006 & -0.017 & -0.0275 & -0.0225 & -0.0065 & -0.0122 & -0.0071\\
0.006 & 0 & -0.0055 & 0 & 0.0075 & -0.0005 & -0.0008 & -0.0002\\
0.017 & 0.0055 & 0 & 0.001 & 0.0262 & 0.0156 & -0.0245 & 0\\
0.0275 & 0 & -0.001 & 0 & -0.0126 & 0.0002 & -0.004 & 0\\
0.0225 & -0.0075 & -0.0262 & 0.0126 & 0 & -0.018 & -0.0095 & -0.029\\
0.0065 & 0.0005 & -0.0156 & -0.0002 & 0.018 & 0 & -0.0035 & 0.0005\\
0.0122 & 0.0008 & 0.0245 & 0.004 & 0.0095 & 0.0035 & 0 & 0.0135\\
0.0071 & 0.0002 & 0 & 0 & 0.029 & -0.0005 & -0.0135 & 0
\end{array}\right)\end{array}\label{eq:new6}
\end{equation}
Density matrices after the distributed  measurement of the phase information for the initial Bell-state-ancilla composite states $|\psi^{-}1\rangle,\,|\phi^{+}0\rangle\,$ and $\,|\phi^{-}1\rangle$
are as follows

\begin{equation}
\begin{array}{lcc}
{\rm Re}\left[\rho^{E}_{|\psi^{-}1\rangle}\right]_{phase} & = & \left(\begin{array}{cccccccc}
0.0371 & 0.0815 & -0.0035 & -0.0152 & -0.0015 & 0.0087 & -0.011 & -0.0627\\
0.0815 & 0.2321 & 0.0492 & -0.0775 & 0.0067 & -0.0095 & -0.0542 & -0.1925\\
-0.0035 & 0.0492 & 0.0201 & -0.0435 & -0.0125 & -0.0012 & 0.001 & 0.0097\\
-0.0152 & -0.0775 & -0.0435 & 0.2331 & 0.0002 & -0.016 & 0.0067 & -0.015\\
-0.0015 & 0.0067 & -0.0125 & 0.00025 & 0.0131 & 0.008 & 0.001 & 0.0015\\
0.0087 & -0.0095 & -0.0012 & -0.016 & 0.008 & 0.0391 & 0.048 & -0.07\\
-0.011 & -0.0542 & 0.001 & 0.0067 & 0.001 & 0.048 & 0.0301 & 0.0175\\
-0.0627 & -0.1925 & 0.0097 & -0.015 & 0.0015 & -0.07 & 0.0175 & 0.3951
\end{array}\right)\end{array}\label{eq:new7}
\end{equation}

\begin{equation}
\begin{array}{lcc}
{\rm Im}\left[\rho^{E}_{|\psi^{-}1\rangle}\right]_{phase} & = & \left(\begin{array}{cccccccc}
0 & -0.0855 & 0.0025 & -0.0622 & -0.0005 & 0.0092 & 0.0072 & 0.0755\\
0.0855 & 0 & -0.0122 & 0.052 & -0.0012 & 0.023 & -0.0217 & 0.03\\
-0.0025 & 0.0122 & 0 & 0.0405 & -0.0022 & -0.0062 & -0.0025 & 0.0042\\
0.0622 & -0.052 & -0.0405 & 0 & 0.002 & 0.0055 & -0.0147 & -0.051\\
0.0005 & 0.0012 & 0.0022 & -0.002 & 0 & -0.0105 & 0 & -0.0482\\
-0.0092 & -0.023 & 0.0062 & -0.0055 & 0.0105 & 0 & 0.0162 & 0.1185\\
-0.0072 & 0.0217 & 0.0025 & 0.0147 & 0 & -0.0162 & 0 & 0.0155\\
-0.0755 & -0.03 & -0.0042 & 0.051 & 0.0482 & -0.1185 & -0.0155 & 0
\end{array}\right)\end{array}\label{eq:new8}
\end{equation}

\begin{equation}
\begin{array}{lcc}
{\rm Re}\left[\rho^{E}_{|\phi^{+}0\rangle}\right]_{phase} & = & \left(\begin{array}{cccccccc}
0.0798 & 0.0135 & 0.007 & -0.0082 & 0.0045 & -0.0022 & 0.0265 & -0.002\\
0.0135 & 0.0108 & 0.0782 & -0.007 & 0.0362 & 0.002 & -0.001 & 0\\
0.007 & 0.0782 & 0.4188 & 0.001 & 0.37 & 0.0427 & -0.016 & 0.0422\\
-0.0082 & -0.007 & 0.001 & 0.0248 & 0.0397 & -0.0025 & -0.0072 & 0.001\\
0.0045 & 0.0362 & 0.37 & 0.0397 & 0.3878 & 0.0535 & 0 & 0.0345\\
-0.0022 & 0.002 & 0.0427 & -0.0025 & 0.0535 & 0.0098 & 0.0415 & -0.0055\\
0.0265 & -0.001 & -0.016 & -0.0072 & 0 & 0.0415 & 0.0468 & -0.0375\\
-0.002 & 0 & 0.0422 & 0.001 & 0.0345 & -0.0055 & -0.0375 & 0.0208
\end{array}\right)\end{array}\label{eq:new9}
\end{equation}

\begin{equation}
\begin{array}{lcc}
{\rm Im}\left[\rho^{E}_{|\phi^{+}0\rangle}\right]_{phase} & = & \left(\begin{array}{cccccccc}
0 & -0.015 & -0.0655 & -0.0565 & -0.0555 & -0.01 & 0.002 & 0.0015\\
0.015 & 0 & -0.023 & 0.0055 & -0.0215i & 0.0005 & 0.0072 & 0.0005\\
0.0655 & 0.023 & 0 & -0.024 & 0.0015 & -0.0455 & 0.045 & 0.0222\\
0.0565 & -0.0055 & 0.024 & 0 & 0.0642 & 0.0005 & 0.0002 & 0.0045\\
0.0555 & 0.0215 & -0.0015 & -0.0642 & 0 & -0.043 & 0.0565 & -0.0242\\
0.01 & -0.0005 & 0.0455 & -0.0005 & 0.043 & 0 & 0.0062 & 0.0045\\
-0.002 & -0.0072 & -0.045 & -0.0002 & -0.0565 & -0.0062 & 0 & 0.0355\\
-0.0015 & -0.0005 & -0.0222 & -0.0045 & 0.0242 & -0.0045 & -0.0355 & 0
\end{array}\right)\end{array}\label{eq:new10}
\end{equation}

\begin{equation}
\begin{array}{ccc}
{\rm Re}\left[\rho^{E}_{|\phi^{-}1\rangle}\right]_{phase} & = & \left(\begin{array}{cccccccc}
0.015 & 0.0145 & -0.0015 & -0.0107 & 0 & 0.0055 & 0.0032 & -0.013\\
0.0145 & 0.05 & 0.0487 & -0.1075 & 0.0205 & 0.0285 & 0.011 & -0.027\\
-0.0015 & 0.0487 & 0.029 & 0.0225 & -0.0017 & -0.0452 & -0.0005 & 0.0077\\
-0.0107 & -0.1075 & 0.0225 & 0.431 & -0.0587 & -0.231 & 0.0007 & 0.0695\\
0 & 0.0205 & -0.0017 & -0.0587 & 0.032 & 0.064 & -0.002 & -0.0137\\
0.0055 & 0.0285 & -0.0452 & -0.231 & 0.064 & 0.23 & 0.0422 & -0.091\\
0.0032 & 0.011 & -0.0005 & 0.0007 & -0.002 & 0.0422 & 0.01 & -0.037\\
-0.013 & -0.027 & 0.0077 & 0.0695 & -0.0137 & -0.091 & -0.037 & 0.203
\end{array}\right)\end{array}\label{eq:new11}
\end{equation}

\begin{equation}
\begin{array}{lcc}
{\rm Im}\left[\rho^{E}_{|\phi^{-}1\rangle}\right]_{phase} & = & \left(\begin{array}{cccccccc}
0 & -0.0085 & 0.0015 & -0.0367 & -0.0527 & -0.0165 & 0.0007 & -0.0106\\
0.0085 & 0 & 0.0157 & 0.14 & -0.004 & -0.0942 & 0.0026 & 0.0335\\
-0.0015 & -0.0157 & 0 & 0.006 & -0.0007 & 0.0351 & -0.0517 & 0.0215\\
0.0367 & -0.14 & -0.006 & 0 & -0.0566 & -0.003 & -0.0025 & -0.0312\\
0.0527 & 0.004 & 0.0007 & 0.0566 & 0 & -0.08 & 0.001 & -0.048\\
0.0165 & 0.0942 & -0.0351 & 0.003 & 0.08 & 0 & -0.004 & 0.026\\
-0.0007 & -0.0026 & 0.0517 & 0.0025 & -0.001 & 0.004 & 0 & 0.0355\\
0.0106 & -0.0335 & -0.0215 & 0.0312 & 0.048 & -0.026 & -0.0355 & 0
\end{array}\right)\end{array}\label{eq:new12}
\end{equation}

Density matrices after the parity measurement for the initial Bell-state-ancilla composite states $|\psi^{-}0\rangle,\,|\phi^{+}1\rangle\,$ and  $\,|\phi^{-}1\rangle$
are

\begin{equation}
\begin{array}{lcc}
{\rm Re}\left[\rho^{E}_{|\psi^{-}0\rangle}\right]_{parity} & = & \left(\begin{array}{cccccccc}
0.4628 & 0.0855 & 0.0165 & 0.0132 & 0.0105 & -0.0127 & -0.334 & -0.0178\\
0.0855 & 0.0038 & 0.0527 & -0.0045 & 0.0007 & 0 & -0.0776 & -0.0032\\
0.0165 & 0.0527 & 0.0308 & -0.0315 & 0.0075 & 0.0018 & 0.01 & 0.0005\\
0.0132 & -0.0045 & -0.0315 & 0.0298 & 0.0106 & -0.0007 & -0.0295 & 0.0025\\
0.0105 & 0.0007 & 0.0075 & 0.0106 & 0.0488 & 0.011 & 0.0105 & 0.0037\\
-0.0127 & 0 & 0.0018 & -0.0007 & 0.011 & 0.0098 & 0.0522 & -0.0055\\
-0.334 & -0.0776 & 0.01 & -0.0295 & 0.0105 & 0.0522 & 0.4008 & 0.0165\\
-0.0178 & -0.0032 & 0.0005 & 0.0025 & 0.0037 & -0.0055 & 0.0165 & 0.0128
\end{array}\right)\end{array}\label{eq:new13}
\end{equation}

\begin{equation}
\begin{array}{lcc}
{\rm Im}\left[\rho^{E}_{|\psi^{-}0\rangle}\right]_{parity} & = & \left(\begin{array}{cccccccc}
0 & -0.0515 & -0.0205 & -0.0662 & -0.011 & 0 & 0.158 & 0.0707\\
0.0515 & 0 & 0.0032 & 0.01 & 0.001 & 0 & -0.0105 & 0\\
0.0205 & -0.0032 & 0 & 0.0485 & -0.01 & -0.0062 & -0.009 & -0.0017\\
0.0662 & -0.01 & -0.0485 & 0 & 0.006 & -0.001 & 0.0057 & -0.0005\\
0.011 & -0.001 & 0.01 & -0.006 & 0 & -0.011 & -0.0175 & -0.041\\
0 & 0 & 0.0062 & 0.001 & 0.011 & 0 & -0.004 & 0.003\\
-0.158 & 0.0105 & 0.009 & -0.0057 & 0.0175 & 0.004 & 0 & -0.027\\
-0.0707 & 0 & 0.0017 & 0.0005 & 0.041 & -0.003 & 0.027 & 0
\end{array}\right)\end{array}\label{eq:new14}
\end{equation}

\begin{equation}
\begin{array}{lcc}
{\rm Re}\left[\rho^{E}_{|\phi^{+}1\rangle}\right]_{parity} & = & \left(\begin{array}{cccccccc}
0.05 & 0.0205 & -0.002 & -0.01725 & -0.002 & -0.0207 & 0.0072 & -0.0052\\
0.0205 & 0.046 & 0.0542 & -0.0935 & 0.0002 & 0.0035 & 0.005 & -0.005\\
-0.002 & 0.0542 & 0.021 & -0.021 & 0.0057 & 0.0267 & 0.003 & -0.014\\
-0.0172 & -0.0935 & -0.021 & 0.427 & 0.092 & 0.189 & -0.018 & 0.0195\\
-0.002 & 0.0002 & 0.0057 & 0.092 & 0.03 & 0.102 & -0.003 & -0.0037\\
-0.0207 & 0.0035 & 0.0267 & 0.189 & 0.102 & 0.239 & 0.0282 & -0.0675\\
0.0072 & 0.005 & 0.003 & -0.018 & -0.003 & 0.0282 & 0.029 & -0.0265\\
-0.0052 & -0.005 & -0.014 & 0.0195 & -0.0037 & -0.0675 & -0.0265 & 0.158
\end{array}\right)\end{array}\label{eq:new15}
\end{equation}

\begin{equation}
\begin{array}{lcc}
{\rm Im}\left[\rho^{E}_{|\phi^{+}1\rangle}\right]_{parity} & = & \left(\begin{array}{cccccccc}
0 & -0.0115 & 0.0045 & -0.054 & -0.0015 & -0.0027 & -0.0027 & -0.0135\\
0.0115 & 0 & 0.004 & 0.0915 & 0.0062 & -0.0085 & -0.0055 & 0.038\\
-0.0045 & -0.004 & 0 & 0 & -0.0037 & -0.044 & -0.002 & 0.0072\\
0.054 & -0.0915 & 0 & 0 & 0.0555 & -0.032 & 0.0132 & -0.0195\\
0.0015 & -0.0062 & 0.0037 & -0.0555 & 0 & -0.0665 & 0.0045 & -0.04\\
0.0027 & 0.0085 & 0.044 & 0.032 & 0.0665 & 0 & 0.007 & 0.0765\\
0.0027 & 0.0055 & 0.002 & -0.0132 & -0.0045 & -0.007 & 0 & 0.0345\\
0.0135 & -0.038 & -0.0072 & 0.0195 & 0.04 & -0.0765 & -0.0345 & 0
\end{array}\right)\end{array}\label{eq:new16}
\end{equation}

\begin{equation}
\begin{array}{lcc}
{\rm Re}\left[\rho^{E}_{|\phi^{-}1\rangle}\right]_{parity} & = & \left(\begin{array}{cccccccc}
0.028 & 0.0155 & 0.002 & -0.032 & 0.0065 & 0.024 & -0.0045 & -0.0041\\
0.0155 & 0.049 & 0.058 & -0.094 & 0.01 & 0.0055 & 0.0073 & 0.006\\
0.002 & 0.058 & 0.015 & -0.022 & -0.0055 & -0.0198 & -0.0005 & 0.0017\\
-0.032 & -0.094 & -0.022 & 0.439 & -0.0838 & -0.2265 & 0.0227 & 0.0195\\
0.0065 & 0.01 & -0.0055 & -0.0838 & 0.026 & 0.1075 & -0.002 & -0.0037\\
0.024 & 0.0055 & -0.0198 & -0.2265 & 0.1075 & 0.256 & 0.0217 & -0.072\\
-0.0045 & 0.0073 & -0.0005 & 0.0227 & -0.002 & 0.0217 & 0.022 & -0.0295\\
-0.0041 & 0.006 & 0.0017 & 0.0195 & -0.0037 & -0.072 & -0.0295 & 0.165
\end{array}\right)\end{array}\label{eq:new17}
\end{equation}

\begin{equation}
\begin{array}{lcc}
{\rm Im}\left[\rho^{E}_{|\phi^{-}1\rangle}\right]_{parity} & = & \left(\begin{array}{cccccccc}
0 & -0.007 & 0.001 & -0.0522 & 0.0025 & 0.002 & 0.0017 & 0\\
0.007 & 0 & 0.0057 & 0.0965 & 0.002 & -0.0025 & 0.0065 & 0.0157\\
-0.001 & -0.0057 & 0 & 0.0035 & 0.0007 & 0.0372 & -0.0005 & 0.0007\\
0.0522 & -0.096 & -0.0035 & 0 & -0.0447 & -0.0072 & -0.0032 & -0.01\\
-0.0025 & -0.002 & -0.0007 & 0.0447 & 0 & -0.0555 & -0.0015 & -0.0282\\
-0.002 & 0.0025 & -0.0372 & 0.0072 & 0.0555 & 0 & 0.0177 & 0.0705\\
-0.0017 & -0.0065 & 0.0005 & 0.0032 & 0.0015 & -0.0177 & 0 & 0.032\\
0 & -0.0157 & -0.0007 & 0.01 & 0.0282 & -0.0705 & -0.032 & 0
\end{array}\right)\end{array}\label{eq:new18}
\end{equation}
\end{widetext}

\end{document}